\documentclass[amsmath,amssymb,pra,reprint,showpacs]{revtex4-1}

\usepackage{graphicx}
\usepackage{dcolumn}

\newcommand{\ket}[1]{|{#1}\rangle}
\newcommand{\bra}[1]{\langle{#1}|}
\newcommand{\prj}[1]{\ket{#1}\bra{#1}}
\newcommand{\ketbra}[2]{\ket{#1}\bra{#2}}

\newcommand{\cA}{{\mathcal A}}
\newcommand{\cS}{{\mathcal S}}

\newcommand{\cT}{{\mathcal T}}

\begin{document}


\title{Light scattering in an optomechanical cavity coupled to a single atom}

\author{Daniel Breyer}
\author{Marc Bienert}%
\affiliation{Theoretische Physik, Universit\"at des Saarlandes, D-66041 Saarbr\"ucken, Germany
}%


\date{\today}

\begin{abstract}
We theoretically analyze the light scattering of an optomechanical cavity which strongly interacts with a single two-level system and couples simultaneously to a mechanical oscillator by radiation forces. The analysis is based on the assumptions that the system is driven at low intensity, and that the mechanical interaction is sufficiently weak, permitting a perturbative treatment. We find quantum interference in the scattering paths, which allows to suppress the Stokes-component of the scattered light. This effect can be exploited to reduce the motional energy of the mechanical oscillator.
\end{abstract}

\maketitle


\section{Introduction}
\label{sec:Intro}

Cavity optomechanics~\cite{om:kippenberg2008,om:marquardt2009} explores the quantum dynamics of electromagnetic and mechanical degrees of freedom coupled by radiation forces. During the last decade, tremendous progress was made in the direction of gaining quantum control over optomechanical systems, including the demonstration of cooling~\cite{om:schliesser2008,om:groeblacher2009,om:chan2011} and strong coupling~\cite{om:schliesser2008b,om:groeblacher2009b} in optical setups. While in typical experiments a laser strongly pumps the optical resonator in order to increase the optomechanical coupling, recent advancements develop towards the regime where the radiative forces of single photons are strong enough to displace the mechanical elements on the order of its ground-state wavepacket fluctuations~\cite{om:eichenfeld2009,gupta2007}. In this regime, the non-linear nature of the optomechanical coupling can manifests itself in phenomenons like photon blockade effect~\cite{om:rabl2011}, creation of non-classical states~\cite{om:bose1997,om:vitali2007,om:mancini1997}, and distinct cooling behaviour becomes apparent~\cite{om:nunnenkamp2012}.

In virtue of the success in atomic physics and cavity quantum electrodynamics~\cite{miller2005} with its well-developed techniques for manipulation of atoms, it is not far to couple atoms with optomechanical cavities. Such hybrid devices can combine the advantages of the different components and provide an additional degree of freedom for controlling the cavity-oscillator system. Strong coupling of the center-of-mass motion of an atom with an optomechanical device was proposed~\cite{om:hammerer2009,*om:wallquist2010}, and the backaction of atoms on a mechanical element and vice versa has been demonstrated recently~\cite{om:camerer2011}. Moreover, there exist first studies about the electronic dynamics of the atom coupled in such setups~\cite{om:chang2009}.
 
In this work we consider an optomechanical cavity interacting with a single two-level system close to the regime where single-photon effects become important. We focus here on the internal degree of freedom of the two-level system, and neglect its center-of-mass motion. Possible realizations of such a setup could be a single atom tightly trapped inside an optical resonator with pendular end-mirror, an artificial atom inside an optomechanical photonic crystal cavity, or a photonic crystal in diamond~\cite{RiedrichMoller2012} coupled to mechanical motion, where a nitrogen-vacancy (NV) center~\cite{aharanovich2011} realizes the two-level system. For such a tri-partite system consisting of a spin degree of freedom, the electromagnetic field of the considered cavity mode, and the mechanical oscillator, we investigate the light scattering at the most fundamental level where a single atom, single photons and single mechanical excitations interact with each other. The treatment here will be in a regime analogous to the Lamb-Dicke regime of atoms in traps~\cite{qo:eschner2003}, where the optomechanical coupling is weak enough to consider it as a perturbation, but still exhibits noticeable effects on the single photon level.

The light scattering at an optomechanical cavity in the single-photon strong coupling regime without atoms was recently investigated~\cite{om:nunnenkamp2012, om:liao2012}. The modulation of the intra-cavity field by the mechanical frequency manifests itself in sidebands appearing in the cavity output spectrum. These features, when used for cooling, give rise to several resonances in the cooling rate~\cite{om:nunnenkamp2012} and lead to non-thermal stationary states in the final stage of the cooling. 

In the work presented here, only the Stokes- and anti-Stokes sidebands are taken into account, in accordance with the assumption of sufficiently weak mechanical interaction. As an additional effect, the presence of the two-level system, strongly coupled to the cavity, changes the optical properties of the composite system drastically. For a fixed cavity interacting with an atom, cavity induced transparency~\cite{qo:rice1996} can emerge, an effect which prohibits the excitation of the atom due to quantum interference. We will identify a similar interference, which involves cavity, mechanical and atomic degrees of freedom. As a main result, we find that in the bad cavity limit, the Stokes-component of the scattered light can thereby be suppressed,  what allows to cool the motion of the mirror ideally close to its ground state, even when the cavity line width is larger than the mechanical oscillation frequency. Moreover, we identify the underlying physical processes of light scattering which include mechanical action.

For a strongly driven optomechanical cavity, a cooling scheme was proposed~\cite{om:genes2011} using a cloud of three-level atoms inside the cavity in order to tailor the cavity's excitation spectrum with the help of electromagnetic induced transparency (EIT) to enhance cooling of the mechanical element. In contrast to this approach, where the atomic cloud acts as an EIT media inside the cavity, here we focus on the strong interaction of elementary excitations of cavity, mechanical object and a single two-level system. Quantum interference effects can not be accredited to the medium, but appear on a more fundamental level in the dynamics involving hybrid quantum states of all degrees of freedom.

The paper is organized as follows: Section~\ref{sec:model} is dedicated to the introduction of the model and its theoretical description. Moreover, the basic assumptions are introduced. In Sec.~\ref{sec:scatter}, the calculation of the scattering rates of light using resolvent theory is performed and light scattering is brought into connection with the dynamics of the mechanical object. The findings are then presented and discussed in Sec.~\ref{sec:results}, while  Sec.~\ref{sec:conclusions} concludes this work. In the appendix we summarize some details of the calculations.

\section{Optomechanical model}
\label{sec:model}

\begin{figure}
 \centerline{\includegraphics[width=6cm]{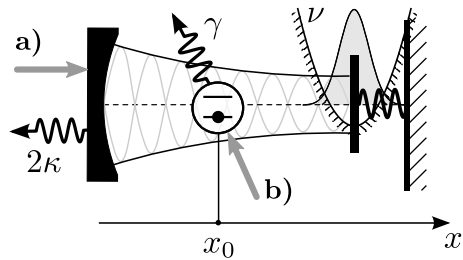}}
 \caption{\label{fig:setup} An optomechanical setup, consisting of a cavity coupled to a mechanical oscillator by radiation pressure, interacts with a single two-level system at fixed position $x_0$ via dipole interaction. We consider two scenarios: \textbf{a)} The cavity is driven;  \textbf{b)} the atom is driven. Both, the atom and the cavity can decay with rates $\gamma$ and $2\kappa$, respectively. The frequency of the mechanical oscillator is $\nu$.}
\end{figure}

The model we investigate consists of a two-level system, fixed at position $x_0$, and coupled to the field of a single mode of an optical resonator by dipole interaction. The resonator's boundary condition can be mechanically varied in a way that the mechanical displacement $x$ changes linearly the frequency of the resonator. The prototype of such an optomechanical setup is a moving end-mirror of a Fabry-Perot cavity, as sketched in Fig.~\ref{fig:setup}. We further assume that the mechanical degree of freedom is confined around its equilibrium position by a harmonic potential with frequency $\nu$. The whole system can be driven either by pumping the cavity \emph{or} the atom with an external laser. We will separately consider both cases in this work. Finally, the coupling of the atom and the cavity to the external radiation field introduces dissipative dynamics. The decay rate of the atomic population is denoted by $\gamma$, and the cavity loss rate by $2\kappa$. 

The total Hamiltonian is given by
\begin{align}
  {\mathcal H } = H + H_{\rm rad}
 \label{eq:Htot}
\end{align}
where $H$ is the Hamiltonian of the optomechanical system of interest, composed of the two-level system, the optical resonator and the mechanical oscillator, and $H_{\rm rad}$ contains the energy of the electromagnetic field outside the cavity, and its interaction with the cavity mode and the atom. 

\subsection{Theoretical description}
\label{sec:hamiltonian}

\subsubsection{Central system: Cavity, two-level system and the mechanical oscillator}

We describe the dynamics of the optomechanical model with the help of the Hamiltonian
\begin{align}
 H = H_{\rm osc} + H_{\rm opt} + F x + W_{\rm L},
 \label{eq:H}
\end{align}
written in a frame rotating with the laser frequency $\omega_{\rm L}$. The first two terms
\begin{align}
  H_{\rm osc} =& \,\hbar \nu [b^\dagger b+\tfrac{1}2]\,,\\
 H_{\rm opt} =& -\hbar \delta \prj{e}-\hbar \Delta a^\dagger a\nonumber\\
 &+\hbar g \left[\ketbra eg a + \ketbra ge a^\dagger\right]\label{eq:Hopt}
\end{align}
account for the energy of the mechanical oscillator with frequency $\nu$ and mass $M$, and the energy of the coupled system of cavity and two-level object, respectively. The annihilation operator $b$ is connected with the oscillator's position and momentum
\begin{align*}
 x = \xi (b + b^\dagger)\,,\quad
 p = \frac{\hbar}{2 i\xi}(b-b^\dagger),
\end{align*}
where $\xi=\sqrt{\hbar/2 M \nu}$ is the width of the oscillator's ground state wavepacket. The two-level system has lower and upper electronic states $\ket g$ and $\ket e$ with transition frequency $\omega_{\rm tls}$, and $a$ is the photon annihilation operator for the cavity's mode with resonance frequency $\omega_{\rm cav}$. The unperturbed quasi-energies of the two levels and the cavity are taken into account by the first and second term in 
 Eq.~\eqref{eq:Hopt}, where the detunings
\begin{align}
 \delta &= \omega_{\rm L}  - \omega_{\rm tls},\\
 \Delta &= \omega_{\rm L} - \omega_{\rm cav}
\end{align}
were defined. In this work we also use the cavity-atom detuning, denoted by
\begin{align}
 \delta_{\rm ca} = \omega_{\rm cav} - \omega_{\rm tls},
\end{align}
such that $\delta = \Delta + \delta_{\rm ca}$.
The electronic and cavity degrees of freedom are coupled in dipole and rotating wave approximation with coupling strength $g=g_0\sin k x_0$ by the Jaynes-Cummings term in Eq.~\eqref{eq:Hopt}, where we assumed a sinusoidal form of the mode along the cavity axis. The maximal coupling $g_0=\vec{\mathcal E}\vec\wp/\hbar$ is proportional to the vacuum electric field $\vec{\mathcal E}$ associated with the mode and to the dipole moment $\vec\wp$ of the transition $\ket{g}\leftrightarrow\ket{e}$.

Small displacements of the mechanical oscillator only slightly change the effective cavity length. This allows to take the mechanical action linearly in the oscillator's displacement $x$ as represented by the term $F x$ in Eq.~\eqref{eq:H}, which has the form
\begin{align}
 F x = -\hbar\chi a^\dagger a (b+b^\dagger)\label{eq:Wrp}
\end{align}
and introduces an interaction between the optical degrees of freedom and the mechanical oscillator.
It accounts for the radiation pressure the electromagnetic field of the driven cavity mode exerts on the mechanical oscillator \cite{om:law1995}. For a Fabry-Perot cavity of length $L$ with pendular end-mirror, $\chi/\xi=\omega_{\rm cav}/L$ corresponds to the radiation force of a single photon over Planck's constant.

Finally, the term $W_{\rm L}$ describes the coherent driving of the system by the pump laser. This part has the form
\begin{align}
  W_{\rm L} &= \hbar\frac{\Omega}2[X + X^\dagger] ,
  \label{eq:WL}
\end{align}
where the operator $X$ stands either for the operator $a$ or $\ketbra ge$ for the case when the cavity or the atom is pumped, respectively. The pump rate is expressed as $\Omega = 2\sqrt{P\kappa/\hbar\omega_{\rm cav}}$ with the power $P$ of the pump laser for the case of a pumped cavity, whereas $\Omega = -\vec\wp\cdot \vec E_0/\hbar$ denotes the Rabi frequency when the laser with electric field amplitude $E_0$ drives the atomic dipole $\vec \wp$. 

\subsubsection{External electromagnetic field}

We complete the theoretical model by defining $H_{\rm rad} =H_{\rm emf}+W_\gamma+ W_\kappa$, and thereby introduce the modes of the electromagnetic field external to the cavity, together with their coupling to the central system. 
The unperturbed energy of these modes in the rotating frame is represented by
\begin{align}
 H_{\rm emf} = \sum_{\vec k, \epsilon} \hbar[\omega_{\vec k}-\omega_{\rm L}] a_{\vec k,\epsilon}^\dagger a_{\vec k,\epsilon}.
\end{align}
The operator $a_{\vec k,\epsilon}$ annihilates a photon in the mode labeled by the wavenumber $\vec k$ and by the polarization index $\epsilon=1,2$, and $\omega_{\vec k}=|\vec k|/c$. The external modes couple to the dipole of the two-level system, 
\begin{align}
 W_\gamma&={\sum_{\vec k, \epsilon}} \,\hbar g^{(\gamma)}_{\vec k,\epsilon} a_{\vec k, \epsilon}^\dagger \ketbra ge + {\rm H.c.}
 \label{eq:Wgamma}
\end{align}
with coupling constant $g^{(\gamma)}_{\vec k,\epsilon}={\mathcal E}_{\vec k}(\vec e_{\vec k,\epsilon}\cdot\vec \wp)e^{i (\vec k\cdot \vec e_x)x}/\hbar$, where ${\mathcal E}_{\vec k}$ denotes the vacuum electric field of mode $\vec k$, $\vec e_{\vec k,\epsilon}$ are the polarization unit vectors and $\vec e_x$ the unit vector in $x$-direction. The cavity losses are modeled by
\begin{align}
 W_\kappa&={\sum_{\vec k, \epsilon}} \,\hbar g^{(\kappa)}_{\vec k,\epsilon} a_{\vec k, \epsilon}^\dagger a + {\rm H.c.}
 \label{eq:Wkappa}
\end{align}
which couples the external field with the cavity mode. For a detailed model which give explicit expressions for $g^{(\kappa)}_{\vec k,\epsilon}$ we refer to the literature~\cite{qo:dutra2000,qob:carmichael:statistical_methods1}. In Eqs.~\eqref{eq:Wgamma} and \eqref{eq:Wkappa} we assume that the coupling constants describe the interaction  of the two-level system and the cavity mode with independent sets of external modes, namely transversal and longitudinal with respect to the cavity axis.

In the Wigner-Weisskopf approach, the couplings, Eqs.~\eqref{eq:Wgamma} and \eqref{eq:Wkappa}, lead to relaxation of the two-level and cavity degree of freedom on a time scale given by the rates
$\gamma = 2\pi {\sum_{\vec k, \epsilon}}|g^{(\gamma)}_{\vec k,\epsilon}|^2\delta(c|\vec k|-\omega_0)$ and $\kappa = \pi {\sum_{\vec k, \epsilon}} |g^{(\kappa)}_{\vec k,\epsilon}|^2\delta(c|\vec k|-\omega_{\rm cav})$ of spontaneous emission and cavity decay, respectively.

\subsection{Assumptions: Small photon numbers and weak mechanical interaction}
\subsubsection{Parameter regime}
\label{sec:pregime}
In this work we apply two main approximations, namely {\it(i)} a weak mechanical interaction, and {\it (ii)} a weak drive of the system. For the first approximation to hold, it is necessary that the optomechanical coupling $\hbar\chi$ is much smaller than the energy difference $\hbar\nu$ between adjacent vibrational levels. Then, only transitions between adjacent vibrational levels are important for the dynamics.
The second approximation ensures that the fundamental mechanical processes associated with single photons are most relevant. We will treat the system's dynamics perturbatively in the small parameters
\begin{subequations}
\begin{align}
 \eta = \chi/\nu&\ll 1\label{eq:condWeakMech},\\
 \langle n\rangle &\ll 1\label{eq:condWeakDrive},
\end{align}
\label{eq:smallpara}
\end{subequations}
where the small cavity photon number $\langle n\rangle$ implies that the resonator is weakly driven, or the pump is sufficiently far detuned from any resonance of the atom-cavity system.

\subsubsection{Relevant basis states of the two-level -- cavity system}

In lowest order perturbation theory, {\it i.e.} when the mechanical and laser interaction vanish, the dynamics of the central system is governed by the Hamiltonian
\begin{align}
 H_0 = H_{\rm opt} + H_{\rm osc}.\label{eq:hamoms}
\end{align}
The corresponding eigenstates are products of Fock states $\ket m$, which diagonalize the mechanical part, and the
eigenstates of the Jaynes-Cummings Hamiltonian $H_{\rm opt}$. Assumption~\eqref{eq:condWeakDrive} implies that only lowest-energy states of the atom-cavity system play a relevant role, {\it i.e.} the states $\{\ket{g,0}, \ket{e,0}, \ket{g,1}\}$. The latter two are not eigenstates of $H_{\rm opt}$, but the lowest-energy dressed states
\begin{subequations}
\begin{align}
 \ket{+} &= \left[\cos\frac{\theta}2 \ket{e,0}+\sin\frac{\theta}2\ket{g,1}\right]\\
 \ket{-} &= \left[\cos\frac{\theta}2 \ket{g,1}-\sin\frac{\theta}2\ket{e,0}\right]
\end{align}
\label{eq:dstates}
\end{subequations}
fulfill $H_{\rm opt}\ket{\pm} = \omega_\pm\ket{\pm}$ 
with eigenfrequencies
\begin{align}
 \omega_{\pm} = -\frac{\delta+\Delta}2\pm\frac{1}{2}\sqrt{(\Delta-\delta)^2+4 g^2},
 \label{eq:dstatesenergy}
\end{align}
where we defined the mixing angle $\theta=\arg \Delta-\delta+2ig$. 
The uncoupled state $\ket{g,0}$ is the ground state of $H_{\rm opt}$ with eigenvalue $\omega_0 = 0$. 

The product states $\ket{g,0; m}=\ket{g,0}\ket{m}$ and $\ket{\pm; m}=\ket{\pm}\ket{m}$ form the unperturbed basis states for the description of the system dynamics. In higher order perturbation theory, they are coupled due to driving with the pump laser, and the mechanical interaction, brought in by the operators $W_{\rm L}$, Eq.~\eqref{eq:WL}, and $Fx$, Eq.~\eqref{eq:Wrp}, respectively. Moreover, the coupling to the external electromagnetic field, described by the terms~\eqref{eq:Wgamma} and \eqref{eq:Wkappa}, leads to instability of the states $\ket{\pm}$, when the relaxation process is treated in Wigner-Weisskopf approximation.

\section{Scattering rates}
\label{sec:scatter}

\subsection{Scattering processes and transition amplitudes}

\begin{figure}
 \includegraphics[width=6cm]{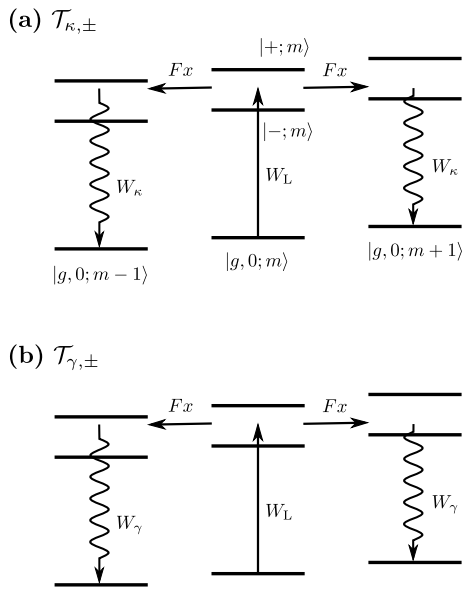}
 \caption{\label{fig:scatterprocess}The Raman scattering processes leading to the transition matrix elements $\cT_{j,\pm}$: The system, initially in the stable state $\ket{g,0;m}$, is excited by the pump interaction $W_{\rm L}$ into a superposition of the dressed state $\ket{\pm;m}$. It passes through these two intermediate states, with an energy defect $E_i-\hbar\omega_\pm-\hbar (m+1/2) \nu$. The mechanical interaction $Fx$ then brings the $\ket{g,1}$-component of the dressed states to an adjacent vibrational level, where the system again passes through the dressed states $\ket{\pm; m\pm 1}$, this time with an energy defect of $E_i-\hbar\omega_\pm-\hbar (m\pm 1+1/2) \nu$. Finally, due to their instability, the dressed states can either decay due to {\bf (a)} cavity loss or {\bf (b)} spontaneous decay, in that way closing the cooling/heating cycle. After many such cycles a stationary occupation of the vibrational levels can be reached, where cooling and heating transitions are balanced.}
\end{figure}

We aim on the analysis of processes, where the single two-level system interacts with a single photon, which -- due to radiation pressure -- couples to the mechanical oscillator. To this end, we apply resolvent theory \cite{qob:cohentannoudji:atom_photon_interaction} and perturbatively calculate the probability amplitudes
\begin{align}
 S_{\rm fi} &= \bra{\rm f}S\ket{\rm i} \nonumber\\
 &=- 2\pi i\, \cT_{\rm fi}(E_{\rm i})\,\delta(E_{\rm i}-E_{\rm f})
 \label{eq:Sfi}
\end{align}
to end in the final state $\ket f$ when the system initially was in state $\ket{\rm i}\neq\ket{\rm f}$
 for conditions~\eqref{eq:smallpara}. The $\delta$-function in Eq.~\eqref{eq:Sfi} ensures that the energies $E_\text{i}$ and $E_\text{f}$ of the initial and final states, determined by the unperturbed part $H_\circ$ of
\begin{align}
 {\mathcal H} = H_\circ + V,
\end{align}
is conserved after the system has passed through the processes described by the transition matrix elements
\begin{align}
 \cT_{\rm fi}(E_{\rm i}) = \bra{\rm f} V\ket{\rm i} + \bra{\rm f} V\frac{1}{E_{\rm i}+i 0^+-{\mathcal H}}V\ket{\rm i}.
 \label{eq:ct}
\end{align}
We arrange the parts
\begin{align}
 H_\circ &=H_{\rm osc} + H_{\rm opt} + H_{\rm emf},\\
 V &= F x + W_{\rm L} + W_\gamma + W_\kappa
\end{align}
of the total Hamiltonian $\mathcal H$, Eq.~\eqref{eq:Htot}, such that the interaction $V$ comprises
the mechanical interaction, the pumping of the system and the coupling to the environment. The strong atom-cavity coupling is contained in $H_\circ$. The rates of cooling and heating transitions can be found by considering  Raman processes with
\begin{align}
 \ket{\rm i} &= \ket{g,0; m}\otimes\ket{\rm vac},\\
 \ket{\rm f} &= \ket{g,0; m\pm 1}\otimes\ket{1_{\vec k, \epsilon}},
\end{align}
whereby $E_{\rm i} = m\hbar\nu +1/2$. Here, $\ket{g,0}$ denotes the stable state of the atom-cavity system, and during the interaction, the vibrational quantum number $m$ changes by one, while finally a photon with wavevector $\vec k$ and polarization index $\epsilon$ is emitted into the external radiation field. The evaluation of the transition matrix, performed in App.~\ref{app:tderivation}, shows that it can be written in the form
\begin{align}
 \cT_{j,\pm} = \alpha_{j,\pm}  \left[{\mathcal S}(E_{\rm i})\; {\mathcal A}_{j,\pm}(E_{\rm i})\right]. \label{eq:tpmform}
\end{align}
We distinguish here between processes where the decay mechanism into the final state is either cavity loss, $j=\kappa$, or atomic fluorescence, $j=\gamma$. In either case, the transition matrix element is proportional to a factor $\cS(E_i)$ representing the probability amplitude for exciting the cavity by the pump. The mechanical interaction is contained in ${\mathcal A}_{j,\pm}(E_{\rm i})$, and this part depends on the decay channel $j$, and on whether a Stokes ($+$) or anti-Stokes ($-$) process is considered. The prefactor $\alpha_{j,\pm}$ is defined in the appendix. With the help of the results~\eqref{eq:tpmform}, we will characterize the dynamics of the motional degree of freedom and of the scattered light. The processes leading to Eq.~\eqref{eq:tpmform} are illustrated in Fig.~\ref{fig:scatterprocess}.

\subsection{Rates for transitions between adjacent vibrational levels}

Light scattering at the optomechanical system leads to transitions between adjacent vibrational levels. Under the assumption~\eqref{eq:condWeakMech} and with the help of the rates
\begin{align}
 R_\pm^{(m)} &\equiv (m+\delta_\pm)A_\pm\nonumber\\ &= \frac{2\pi}{\hbar}\sum_{\rm f}|\cT_{\rm fi}|^2\delta(E_{\rm i}-E_{\rm f}),
 \label{eq:Apm}
\end{align}
corresponding to the Raman processes which we described using the matrix elements \eqref{eq:tpmform}, the change of population~\cite{qo:stenholm1986}
\begin{align}
 \frac{d}{dt}p_m = &-[mA_-+(m+1)A_+] p_m\nonumber\\
                   &+(m+1) A_- p_{m+1}+m A_+ p_{m-1}
\label{eq:rateq}
\end{align}
in the vibrational level $m$ can be formulated ($\delta_+=1$ and otherwise zero). From Eq.~\eqref{eq:rateq} follows that the mean phonon number $\langle m\rangle =\sum_m p_m m$ changes with the rate
\begin{align}
 \Gamma = A_--A_+.
 \label{eq:crate}
\end{align}
For $A_--A_+>0$ scattering to energetically lower vibrational levels prevails, and the populations converge to a stationary thermal distribution $p_m\propto (A_+/A_-)^m$ with mean occupation number 
\begin{equation}
 \langle m\rangle_\infty = \frac{A_+}{\Gamma}.
 \label{eq:meanm}
\end{equation}
While Eq.~\eqref{eq:rateq} characterizes the dynamics of the mechanical element, the scattering rate of photons into the Stokes and Anti-Stokes components is given by Eq.~\eqref{eq:Apm}. In the stationary state, an average over all levels $m$ yields
\begin{align}
R_\pm=(\langle m\rangle_\infty + \delta_\pm)A_\pm. 
\label{eq:Rpm}
\end{align}
Then, the number of photons scattered per unit time into the Stokes and Anti-Stokes sidebands is equal, in accordance with the detailed balance condition. In the non-stationary case, the momentary value of $\langle m\rangle$ enters Eq.~\eqref{eq:Rpm}, which tends to $\langle m\rangle_\infty$ on a time scale given by $1/\Gamma$, and the components are not of equal strength as cooling is ongoing ($\Gamma>0$).

\subsection{Transition rates of heating and cooling}

In this section we present the explicit form of the rates $A_\pm$. The evaluation of Eq.~\eqref{eq:Apm} yields
\begin{align}
 A_\pm = |\cS|^2\left[\,2\kappa|\,\cA_{\kappa,\pm}|^2 +  \gamma\,|\cA_{\gamma,\pm}|^2\,\right]
 \label{eq:Apmres}
\end{align}
with 
\begin{align}
 |\,\cA_{\kappa,\pm}|^2 &=  \chi^2\frac{(\delta\mp\nu)^2+{\gamma^2/4}}{D(\mp\nu)}\label{eq:Ak}\\
 |\,\cA_{\gamma,\pm}|^2 &=  \chi^2\frac{g^2}{D(\mp\nu)}\label{eq:Ag}.
\end{align}
The heating and cooling rates $A_\pm$ are proportional to the excitation probability of the cavity $|\cS|^2$, and consist of two terms which scale with the decay rates of cavity and atomic losses and describe the mechanical processes induced by radiation pressure on the mechanical element, whereby the inserted photon finally leaks out of the cavity or is spontaneously emitted by the atom.

In Eqs.~\eqref{eq:Ak} and \eqref{eq:Ag}, we introduced the denominator
\begin{align}
 D(\upsilon)=&\left[(\delta+\upsilon)(\Delta+\upsilon)-g^2-\frac{\gamma\kappa}2\right]^2\nonumber\\
           &\quad\quad\quad+\left[\kappa(\delta+\upsilon)+\frac{\gamma}2(\Delta+\upsilon)
           \right]^2    
\end{align}
giving rise to resonances for $\omega_\pm-\upsilon=0$, as long as $\kappa, \gamma$ are sufficiently small such that the system is not overdamped. At this condition, the pump is detuned from one of the dressed states~\eqref{eq:dstates} by $\upsilon$. 

It remains to present the excitation probabilities for the two different pump schemes defined by Eq.~\eqref{eq:WL}. One finds
\begin{align}
 |\cS|^2 &= \frac{\Omega^2}4\frac{\delta^2+\gamma^2/4}{D(0)}&\text{(cavity pumped)},
 \label{eq:Sc}\\
 |\cS|^2 &= \frac{\Omega^2}4\frac{g^2}{D(0)}&\text{(TLS pumped)} .
 \label{eq:Sa}
\end{align}
Those have the same functional dependency as $\cA_{j, \pm}$, Eqs.~\eqref{eq:Ak} and \eqref{eq:Ag}, however, without the shifts by the oscillation frequency $\mp\nu$, since no mechanical interaction is involved in the excitation by the pump.

\section{Discussion}
\label{sec:results}

\begin{figure}
 \includegraphics[width=8cm]{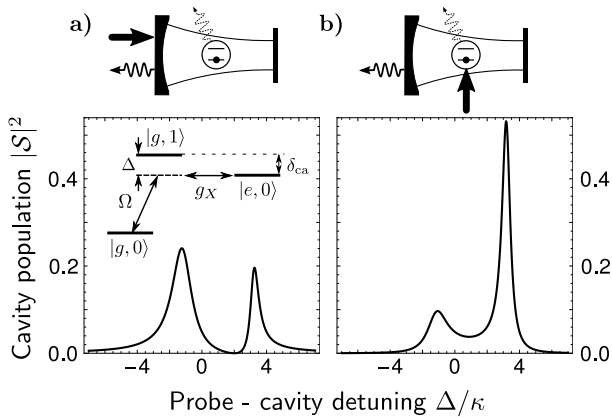}
 \caption{\label{fig:exc} Population in the single-photon state of the cavity when either {\bf a)} the fixed cavity is directly driven, or {\bf b)} indirectly via the atom. The two peaks correspond to the dressed states $\ket\pm$ of the atom-cavity system. In {\bf a)} a dark resonance appears at $\delta=\Delta+\delta_{\rm ca}=0$, {\it i.e.} when the pump is resonant with the atomic transition. This effect is due to coherent population trapping in the $\Lambda$-type configuration shown in the inset. (Parameters: $\gamma=0.1\kappa$, $g=2\kappa$, $\delta_{\rm ca}=-2\kappa$, $\Omega=\kappa$.)}
\end{figure}

All the transition rates we found in the last section are proportional to the cavity's excitation probability $|\mathcal S|^2$ for the two considered options of driving the system. We start the discussion by pointing out its characteristics and peculiar features. A plot of $|\mathcal S|^2$ is shown in Fig.~\ref{fig:exc} for both driving schemes. The case of a driven cavity is shown in part a) of the figure: The cavity can be best excited, when the laser is tuned close to one of the two dressed state frequencies $\omega_\pm$, where the curve shows resonances. For $\gamma\ll g$, a distinctive feature emerges in the population of the cavity: Around $\Delta = -\delta_{\rm ca}$ the curve shows a Fano like profile and drops down to almost zero at $\Delta=-\delta_{\rm ca}$ due to an interference effect similar to cavity induced transparency~\cite{Rice1996}, where the state $\ket{g,1}$ can not be populated because of destructive quantum interference in the excitation paths, see inset of the figure.
The cavity excitation as a function of the pump frequency in Fig.~\ref{fig:exc}b), does not show this feature, but only the two resonances connected with the dressed states $\ket{\pm}$. 

\begin{figure*}
 \centerline{\includegraphics[width=12cm]{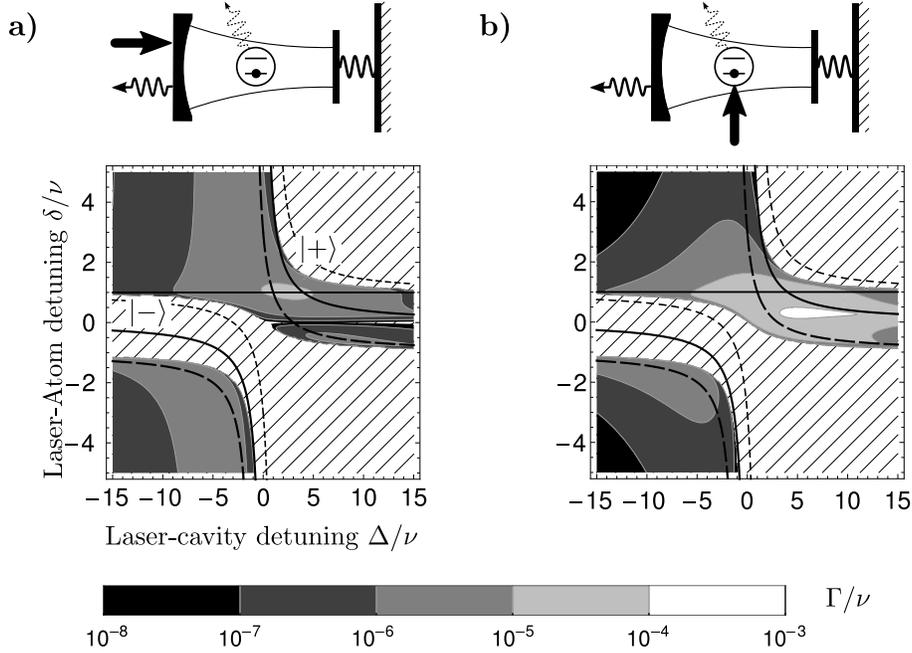}}
 \caption{\label{fig:coolingrate} Cooling rate $\Gamma$ as a function of the laser-cavity detuning $\Delta$ and the laser-atom detuning $\delta$ for the cases of \textbf{a)} driven cavity and \textbf{b)} driven atom. Bright shaded regions represent large cooling rates, in the hatched area the rate is negative what corresponds to heating the mechanical oscillator. Shown is also the resonance condition $\omega_\pm(\Delta,\delta)=0,-\nu,\nu$, where the laser is in resonance with the dressed state $\ket\pm$ (solid curve), or with the blue (dashed) or red (broken) sidebands of the dressed states, respectively.
  (Parameters: $\kappa = 7\nu, \gamma = 0.05\nu, \Omega = \nu, \chi=\nu/10, g=2\nu$).}
\end{figure*}

With these results we now turn to the influence of light scattering on the mechanical degree of freedom. In Fig.~\ref{fig:coolingrate} we show the rate $\Gamma$, Eq.~\eqref{eq:crate}, at which the mechanical degree of freedom approaches its steady state when $\Gamma>0$. It is plotted as a function of the pump-cavity detuning $\Delta$ and the pump-atom detuning $\delta$ for the two pump configurations. In the plots, brighter areas show larger cooling rates, whereby in the hatched area, heating transitions dominate over cooling transitions ($\Gamma<0$) and no steady state can be found within the validity of the theory presented here. The parameters are chosen such that the cavity loss is the dominating dissipative mechanism: $\kappa\gg\gamma$. Then, according to Eq.~\eqref{eq:Ak}, for $\delta = \nu\gg\gamma$ one finds $A_+\approx 0$: A similar interference effect as found in Fig.~\ref{fig:exc}a) manifests itself, this time, however, between two different scattering paths involving mechanical transitions. As in EIT cooling~\cite{qo:morigi2000,*qo:morigi2003}, or cooling schemes of trapped atoms in cavities~\cite{qo:zippilli2005a,*qo:zippilli2005,bienert:2012}, such a behaviour can be exploited in order to suppress undesired transitions which lead to heating.
Indeed, in Fig.~\ref{fig:coolingrate}a) at $\delta=\nu$ a signature of suppressed heating can be observed. It becomes most significant at the crossing with the black curves, which symbolize the resonance condition of the pump with the dressed states $\ket\pm$ (solid curve), or with the dressed states belonging to the adjacent lower vibrational level shifted by the frequency $-\nu$ (broken curve). 

When following the curves of the dressed state resonances towards large values of $\Delta$, the states $\ket\pm\sim\ket{e,0}$ become atom-like and exhibit a narrow linewidth. When the dressed state's linewidth becomes much smaller than the trap frequency $\nu$, the laser can address the red sideband of this resonance in order to preferably scatter Anti-Stokes photons. This sideband cooling at the dressed state can be seen in the figure, where the cooling rate becomes large around the broken curve at large $|\Delta|$. The dressed states are photon-like $\sim\ket{g,1}$ around $\Delta = 0$ and for large values of $|\delta|$. For the parameters chosen here, the states have a broad linewidth $\gamma_\pm$, explaining the extended cooling areas at moderate cooling rates on the red side of the dressed states resonances. 

Increased cooling rates are also found along the solid curve for $\Delta>0$, reflecting the enhanced cavity excitation probability found in Fig.~\ref{fig:exc}a), when a dressed state is addressed by the pump. Similarly, the strong reduction of $\Gamma$ at $\delta = 0$ can be explained by the excitation probability, Eq.~\eqref{eq:Sc}, namely when induced transparency hinders a significant excitation of the cavity.

In part b) of Fig.~\ref{fig:coolingrate} the cooling rate is shown for the same parameters, however, when the system is driven via the two-level system. The highest rates $\Gamma$ are found in a larger area around the atomic resonance $\delta = 0$ and for $\Delta>0$, which extends into the parameter region of suppressed heating and sideband cooling at $\ket+$, such that the different effects can not be clearly distinguished for the chosen parameters. We observe generally higher cooling rates when the atom is driven, especially when the pump is in resonance $\delta=0$.

\begin{figure}
 \includegraphics[width=6.5cm]{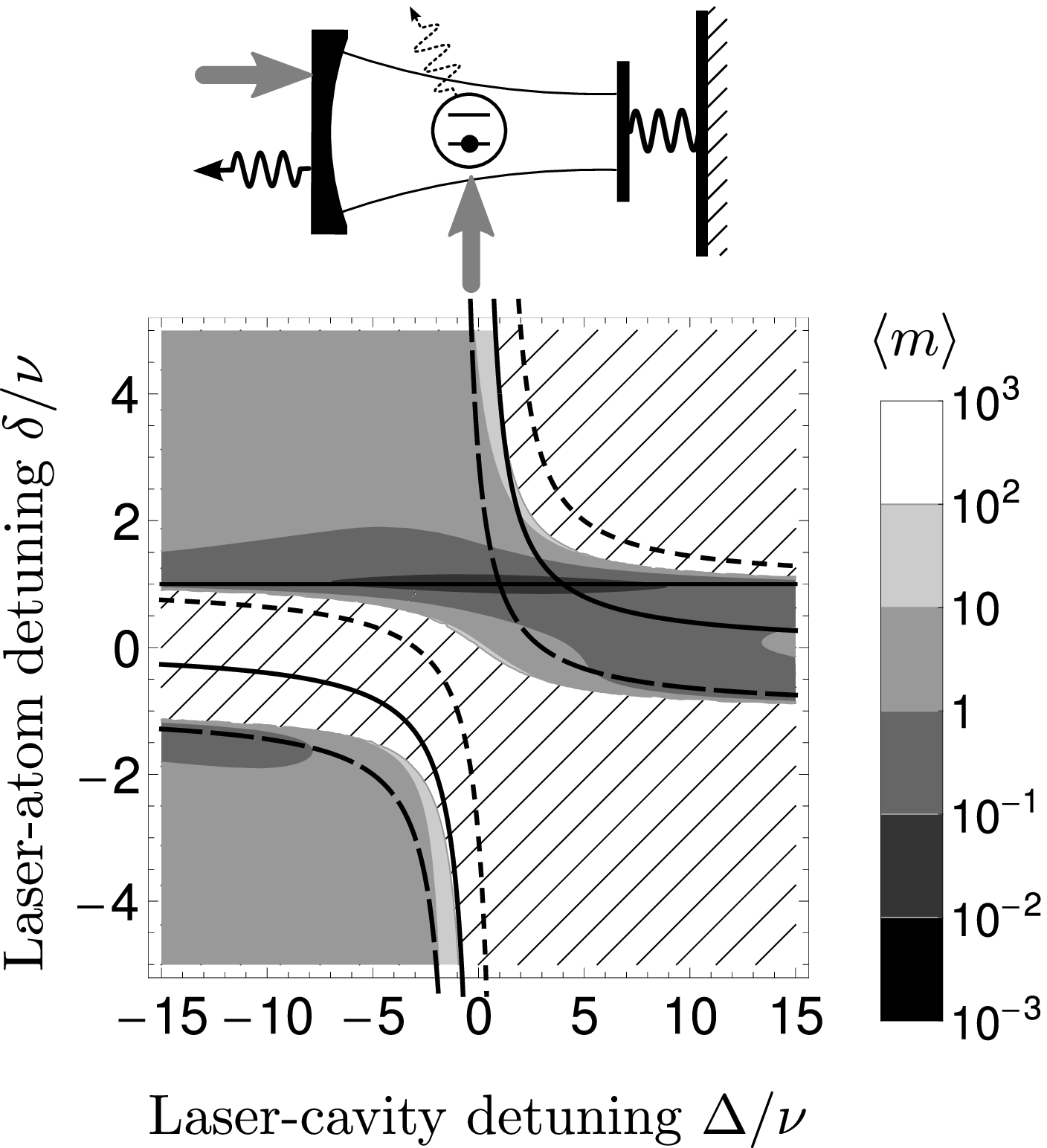}
 \caption{\label{fig:temperature} The mean vibrational occupation number $\langle m\rangle$ as a function of the laser-cavity detuning $\Delta$ and the laser-atom detuning $\delta$, being equal for the driven atom and the driven cavity. Darker shading represents lower temperatures, in the hatched regions the theory does not provide a stationary state. The curves are connected with the dressed states' frequencies and their sidebands as explained in Fig.~\ref{fig:coolingrate}.}
\end{figure}

In contrast to the cooling rate $\Gamma$, the mean vibrational occupation number $\langle m\rangle_\infty$ does not depend on the pump configuration, what becomes clear from its definition~\eqref{eq:meanm} and the form~\eqref{eq:Apmres} of the rates $A_\pm$. Fig.~\ref{fig:temperature} shows $\langle m\rangle_\infty$ for the same parameters as in Fig.~\ref{fig:coolingrate}. Dark shading marks low phonon numbers and hence low temperatures. The lowest occupation numbers, which can lead to ground state cooling, are situated around the resonance condition $\delta = \nu$, where the Stokes-scattering $A_+\approx 0$ is strongly reduced. Apart from this area, low phonon numbers can be reached by sideband cooling at the dressed states when the linewidth becomes $~\sim\gamma\ll\nu$ for a sufficiently far detuned laser from the cavity resonance (large $|\Delta|$). For asymptotic values of $|\delta|$, when the dressed states have a large linewidth $\sim\kappa$, Doppler-like cooling leads to the broad cooling areas at moderate temperatures on the red side of the dressed state curves. Note that for $\delta = -\nu$, quantum interference leads to the opposite effect $A_-\approx 0$, and hence heating dominates, just as along the dashed curves which show the resonances of the laser with the blue sideband of the dressed states.

For $\kappa\gg\gamma$ and $\delta=\nu$ and the optimal value $\Delta_\text{opt}=(\kappa\gamma/2 + g^2)/2\nu-\nu$ one finds a mean vibrational occupation number
\begin{align}
 \langle m\rangle_\infty = \frac{1}C + \left(\frac{\gamma}{4\nu}\right)^2
 \label{eq:minmeanm}
\end{align}
and 
\begin{align}
 \Gamma = \frac{2\chi^2}{\kappa}|\cS|^2
 \label{eq:Gammaopt}
\end{align}
when evaluating Eq.~\eqref{eq:meanm} together with Eqs.~\eqref{eq:Apmres}. The final temperature is therefore determined by the single atom cooperativity $C=2 g^2/\kappa\gamma$, which measures the strength of the coherent cavity coupling, and the atomic noise, described by the second term in Eq.~\eqref{eq:minmeanm}.
When $\delta \to \infty$, or $\varphi=0$ meaning that the atom is placed at a node of the cavity, the presence of the two-level system becomes irrelevant, and one can perform sideband or Doppler cooling at the cavity resonance depending on if $\kappa\ll\nu$ or vice versa. This is similar to the typical laser cooling schemes~\cite{om:wilsonrae2007,om:marquardt2007} of optomechanical devices, but in the low intensity limit.
For $\Delta\to\infty$, one has $\Gamma\to 0$, since the cavity can only hardly be excited. Asymptotically $\Gamma\sim (2 g/\Delta)^2\chi^2/\gamma|\cS|^2$, and the minimal temperature is the same as in Eq.~\eqref{eq:minmeanm}. 

\begin{figure}
 \includegraphics[width=\columnwidth]{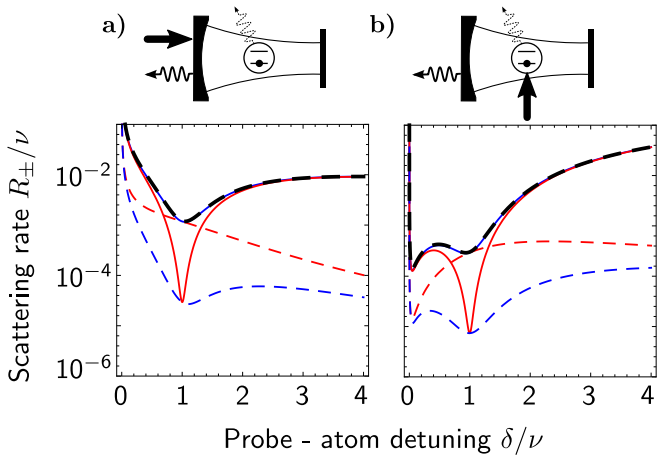}
 \caption{\label{fig:sb} Scattering rate of photons into the Stokes (red) and anti-Stokes (blue) component of the cavity output (solid lines) and fluorescence signal (dashed lines) for $\kappa\gg\gamma$ and $\Delta =0$ at steady state. The black dashed line being the sum of the blue or red curves gives the total rate $R_\pm$ of scattered photons into the motional sidebands, which in steady state are equal for heating and cooling transitions. In {\bf a)} the cavity is driven, {\bf b)} shows the case of a driven atom. (Parameters: $\kappa=5\nu$, $\gamma=0.1\nu$, other parameters like in Fig.~\ref{fig:coolingrate}.)}
\end{figure}

Let us now consider the properties of the scattered light. Scattered photons, which triggered a transition between the mechanical states are scattered into Stokes- and Anti-Stokes sidebands around the carrier frequency at $\omega_{\rm L}\pm\nu$. The sidebands can be observed in both the cavity output and the atomic fluorescence spectrum. In Fig.~\ref{fig:sb} we show the corresponding rates $R_{\gamma;\pm}$ and $R_{\kappa;\pm}$, which add up to the total rate $R_\pm$, Eq.~\eqref{eq:Rpm}, (black curve), as a function of $\delta$ in the stationary state. Again, we focus on the case $\kappa\gg\gamma$, and indeed, for most values of $\delta$, the scattering along the cavity (solid curves) dominate. At $\delta = \nu$, however, the situation changes: Quantum interference suppresses the scattering of Stokes photons through the cavity, the red solid curve drops down. Nevertheless, in the stationary state, the rates $R_\pm$ have to be equal. Hence, the corresponding blue sideband also decreases, until it is compensated by the Stokes scattering at the atom (dashed red curve). Small values of $R_\pm$ indicate low temperatures (or small cooling rates) as $R_\pm=\langle m\rangle_\infty[\langle m\rangle_\infty +1]\Gamma$.

In the model presented here, we did not take into account the intrinsic damping of the mechanical oscillator introduced by its coupling to the environment. A straightforward extension, which is justified when relaxation takes place on a much slower time scale than the cooling processes $\sim1/\Gamma$, is possible by incorporating the thermalization rates in a redefinition of 
\begin{align}
 A'_+ &= A_+ + m_{\rm th}\gamma_{\rm th},\\
 A'_- &= A_- + (m_{\rm th}+1)\,\gamma_{\rm th}.
\end{align}
Here, $m_{\rm th}$ is the equilibrium occupation number of the unperturbed oscillator at temperature $T$ and $\gamma_{\rm th}$ is the mechanical damping rate. Such a description requires large values of the mechanical $Q$-factor. The final temperature due to the atom-assisted laser cooling is then given by
\begin{align}
 \langle m\rangle'_\infty = \frac{1}{\gamma_{\rm th}+\Gamma} \left(\Gamma\,\langle m\rangle_\infty + \gamma_{\rm th} \,m_{\rm th}\right).
\end{align}

With the optimal value of $\Gamma$ from Eq.~\eqref{eq:Gammaopt}, one has $\Gamma_{\rm opt}/\gamma_{\rm th} = 2\eta^2 |S|^2 Q\nu/\kappa$. 
In a cryogenic environment and for $Q\gtrsim10000$, realizable with current~\cite{om:safavi2012, hofer2010} or near-future technology~\cite{ni2012}, notable signatures of the quantum interference involving the atomic, photonic and mechanical quantum degrees of freedom should be detectable in the cooling dynamics, in such a way demonstrating the coherent dynamics of the composite system.

\section{Conclusions}
\label{sec:conclusions}

In this article we analyzed the light scattering in an optomechanical system, consisting of an optical resonator coupled to a mechanical oscillator, where the electromagnetic field inside the cavity interacts strongly with a single two-level system. The analysis was performed in the limit of weak excitation of the system by a pump laser, and for sufficiently weak mechanical interaction. We found that the presence of the atom changes the optical properties of the cavity drastically, and gives rise to interference effects in the scattering processes, allowing to strongly suppress the Stokes scattered light when the decay of the system predominantly takes place due to cavity losses. Then, ideally, the oscillator can be cooled down close to its ground state.
Apart from the interference effect the model also predicts cooling regions when the pump is detuned to the red sideband of the relevant dressed states of the two-level--cavity system, what leads, depending on the effective linewidth of the resonances, to Doppler or sideband cooling like schemes.

In our approach, we took into account the radiation pressure between light and mechanical oscillator, but ignored mechanical corrections in the atom-light interaction~\cite{om:chang2009} which modifies the Jaynes-Cummings term due to the motion of the mechanical element. Such a term $\propto g_0 \xi/L$ is typically a small correction, and is only expected to become important for ultra strong atom-cavity coupling, which might arise in certain setups, for example in solid states devices~\cite{gunter2009}, or when simultaneously several atoms are excited by a single photon~\cite{meiser2006}. Such an interaction leads to a description analogous to cooling the center-of-mass motion of an \emph{atom} by cavity cooling in the low excitation limit~\cite{bienert2012,qo:zippilli2005a,*qo:zippilli2005}. Also in this case, one finds interference effects which can strongly influence the cooling. If both couplings are equally strong, the corresponding cooling processes can interfere and give rise to a complicated cooling dynamics of the mechanical oscillator resulting from the interplay of radiation pressure coupling and the oscillator-position dependent coupling to the atom.

The analysis of the light scattering provided in this work can be relevant for future hybrid quantum systems, where a spin-like system is coupled to an optomechanical device, which might be realized in solid state system or for NV-centers coupled to a photonic crystal cavity which is sensible to vibrations of the structured diamond bulk.

\begin{appendix}

\section{Evaluation of the transition matrix elements}
\label{app:tderivation}

In this appendix we perform the evaluation of the transition matrix elements, Eq.~\eqref{eq:ct}, under the assumptions~\eqref{eq:smallpara}, what finally leads to the form given in Eq.~\eqref{eq:tpmform}. The emission of the photon at the end of the scattering process can be caused either by cavity losses or by spontaneous emission, and hence, we can distinguish between two transition amplitudes
\begin{align}
 \cT_{j,\pm} (E_{\rm i})
 &\approx \bra{\rm f}W_j \,\frac{1}{E_{\rm i}-H_{\rm eff}}\, F x \,\frac{1}{E_{\rm i}-H_{\rm eff}}W_{\rm L}\ket{\rm i}
 \label{eq:tpm}
\end{align}
with $j=\kappa,\gamma$. For writing down Eq.~\eqref{eq:tpm}, we used the definitions Eq.~\eqref{eq:Wrp},  Eq.~\eqref{eq:Wgamma} and Eq.~\eqref{eq:Wkappa}. Moreover, we already expanded the transition matrix elements up to first order in the smallness parameters, Eqs.~\eqref{eq:smallpara}, and introduced the effective Hamiltonian
$H_{\rm eff} = H_\circ - i\hbar\frac{\gamma}{2}\ketbra ee - i\hbar\kappa a^\dagger a$,
which expresses the radiative instability of the atomic and cavity's excited states~\cite{qob:cohentannoudji:atom_photon_interaction}. The radiation pressure, described by the term $Fx\propto a^\dagger a$ in Eq.~\eqref{eq:tpm}, only contributes for a finite number of photons in the cavity. Consequently, one only finds a contribution to scattering, when the transition passes through the state $\ket{\rm i'}=\ket{g,1;m}\otimes\ket{\text{vac}}$, and thus the transition matrix elements
\begin{align}
  \cT_{j,\pm} (E_{\rm i}) = \alpha_{j,\pm}  \left[{\mathcal S}(E_{\rm i})\; {\mathcal A}_{j,\pm}(E_{\rm i})\right] \label{eq:tpm2}
\end{align}
are a product of the amplitudes
\begin{align}
   {\mathcal S}(E_{\rm i}) &= \phantom{-}\bra{\rm i'}\frac{1}{E_{\rm i}-H_{\rm eff}}W_L\ket{\rm i}\label{eq:S},\\
  {\mathcal A}_{j,\pm}(E_{\rm i}) &= -\bra{f}W_j\frac{1}{E_{\rm i}-H_{\rm eff}} x\ket{\rm i'}\,\frac{\chi}{\xi\alpha_{j,\pm}}.
\end{align}
Here, $\mathcal S(E_i)$ is the probability amplitude to excite the cavity by the laser, while ${\mathcal A}_{j,\pm}(E_{\rm i})$ is proportional to the probability amplitude of a decay either due to spontaneous emission ($j=\gamma$) or cavity loss ($j=\kappa$), when the cavity was initially excited and the vibrational quantum number changed by one. In Eq.~\eqref{eq:tpm2} we also introduced the factor $\alpha_{j,\pm}= \sqrt{m+\delta_\pm}\hbar g_{\vec k,\epsilon}^{(j)}$ whereby the symbol $\delta_\pm$ gives unity for Stokes scattering ($+$-case) and otherwise vanishes.
\end{appendix}

\begin{acknowledgments}
We would like to thank Giovanna Morigi for fruitful discussions and enduring encouragement. Moreover, we are grateful to Mauricio Torres for helpful comments.
\end{acknowledgments}


%

\end{document}